\RequirePackage{amsmath}
\documentclass[10pt]{iopart}

\usepackage{bm}
\usepackage{graphicx}
\usepackage{cite}
\usepackage{siunitx}

\begin{document}

\title[Bloch oscillations in optical waveguide arrays with long-range coupling]%
{Exact Bloch oscillations in optical waveguide arrays with arbitrary long-range coupling}

\author{R. Arjona, E. D\'{\i}az and F. Dom\'{i}nguez-Adame}

\address{GISC, Departamento de F\'{\i}sica de Materiales, Universidad Complutense, E-28040 Madrid, Spain}

\ead{elenadg@fis.ucm.es}

\date{\today}

\begin{abstract}
 
We find the exact Bloch oscillations in zigzag arrays of curved optical waveguides under the influence of arbitrary long-range coupling. The curvature induces a linear transverse potential gradient in the equations of the light evolution. In the case of arrays with second-order coupling, steady states can be obtained as linear combinations of Bessel functions of integer index. The corresponding eigenvalues are equally spaced and form the well-known Wannier-Stark ladder, the spacing being independent of the second-order coupling. We also solve exactly the wave packet dynamics and compare it with experimental results. Accordingly we find that a broad optical pulse performs Bloch oscillations. Frequency doubling of the fundamental Bloch frequency sets up at finite values of the second-order coupling. On the contrary when a single waveguide is initially excited, a breathing mode is activated with no signature of Bloch oscillations. We present a generalization of our results to waveguide arrays subject to long-range coupling. In the general case the centroid of the wave packet shows the occurrence of multiples of the Bloch frequency up to the order of the interaction.

\end{abstract}


\submitto{\JO}

\maketitle

\ioptwocol

\section{Introduction}   \label{sec:intro}

Classical and quantum waves in tilted periodic potentials may undergo coherent oscillations, in real and in $k$ space, known as Bloch oscillations~(BOs)~\cite{Bloch29,Zener34}. Since their discovery, BOs became a paradigm of wave coherence in many physical systems. They were observed for the first time as coherent oscillations of electrons in semiconductor superlattices~\cite{Feldmann92,Leo92} (see Ref.~\cite{Leo98} for a review). These oscillations are related to the wave dynamics and thus they can be observed in almost any coherent motion of classical and quantum waves in tilted periodic media. In fact, they were later detected as a periodic time evolution of ensembles of ultracold atoms~\cite{BenDahan96,Wilkinson96} and Bose-Einstein condensates~\cite{Anderson98} in optical lattices. BOs also have their counterpart in acoustics~\cite{Sanchis07} and optics~\cite{Lenz99,Pertsch99,Sapienza03,Lenz03,Longhi06,Longhi09}. Moreover, BOs are interesting not only from a fundamental point of view but pave the way for a number of applications. In this regard, Grenzer \emph{et al.} reported that a semiconductor superlattice structure, coupled by an antenna to a microwave oscillator based on BOs, represents an efficient source of microwave radiation~\cite{Grenzer95}.

BOs last until lattice imperfections or non-linear interactions cause random scattering of different $k$-components of the wave packet, thus broadening its momentum distribution and destroying coherent oscillations in real space. For instance, in Bose-Einstein condensates, the atom-atom interaction gives rise to well-known dynamical instabilities which can destroy the coherence required to sustain stable BOs~\cite{Trombettoni01,Roati04,Fallani04}. Therefore, disorder and interactions present a challenge to observe BOs in experiments. Although decoherence effects cannot be removed completely in matter waves, their impact can be substantially reduced in order to achieve long-living BOs. In this context, Peyrard-Bishop-Holstein polarons in biased DNA molecules can display coherent oscillations in spite of the non-linearity due to the electron-lattice interaction and the disorder of the molecule~\cite{Diaz08,Diaz09}. In Bose-Einstein condensates the non-linear interaction of the Gross-Pitaevskii equation harmonically modulated in time enhances the survival of BOs~\cite{Gaul09,Diaz10}. Moreover, one-dimensional systems with long-range correlations in the disorder show clear signatures of a Bloch-like oscillatory motion~\cite{Adame03}. 

One of the major advantages offered by optical waveguide arrays is the possibility to directly observe the wave packet dynamics in real space with negligible losses~\cite{Longhi06}. With this in mind, Wang \emph{et al.} studied nontrivial BOs in optical waveguide arrays with second-order coupling, i.e. next-nearest neighbor coupling~\cite{Wang10}. They numerically found that a double turning-back occurs when the beam approaches the band edge because BOs almost directly reflect the band structure. Later, experiments verified this remarkable result in zigzag arrays of curved waveguides~\cite{Dreisow11}. Motivated by the occurrence of second-order harmonics of the fundamental Bloch frequency $\omega_B$ when second-order coupling between the waveguides of an array is taken into account, in this work we reexamine the dynamics of wave packets in optical waveguide arrays subject to a linear potential and arbitrary long-range coupling. In case of second-order coupling, we show analytically that doubling of the fundamental Bloch frequency emerges in spite of the fact that the eigenvalues are equally spaced, named Wannier-Stark ladder. Interestingly, the spacing becomes independent of the second-order coupling and thus the Wannier-Stark ladder carries no information of the magnitude of the coupling. Most importantly, we generalize the results to arrays with long-range coupling and find exactly the dynamics of the centroid of the wave packet that displays multiples of the Bloch frequency up to the order of the interaction.

\section{Steady states under second-order coupling}   \label{sec:steady_states}

\begin{figure}[htb]
\centerline{\includegraphics[width=0.95\linewidth]{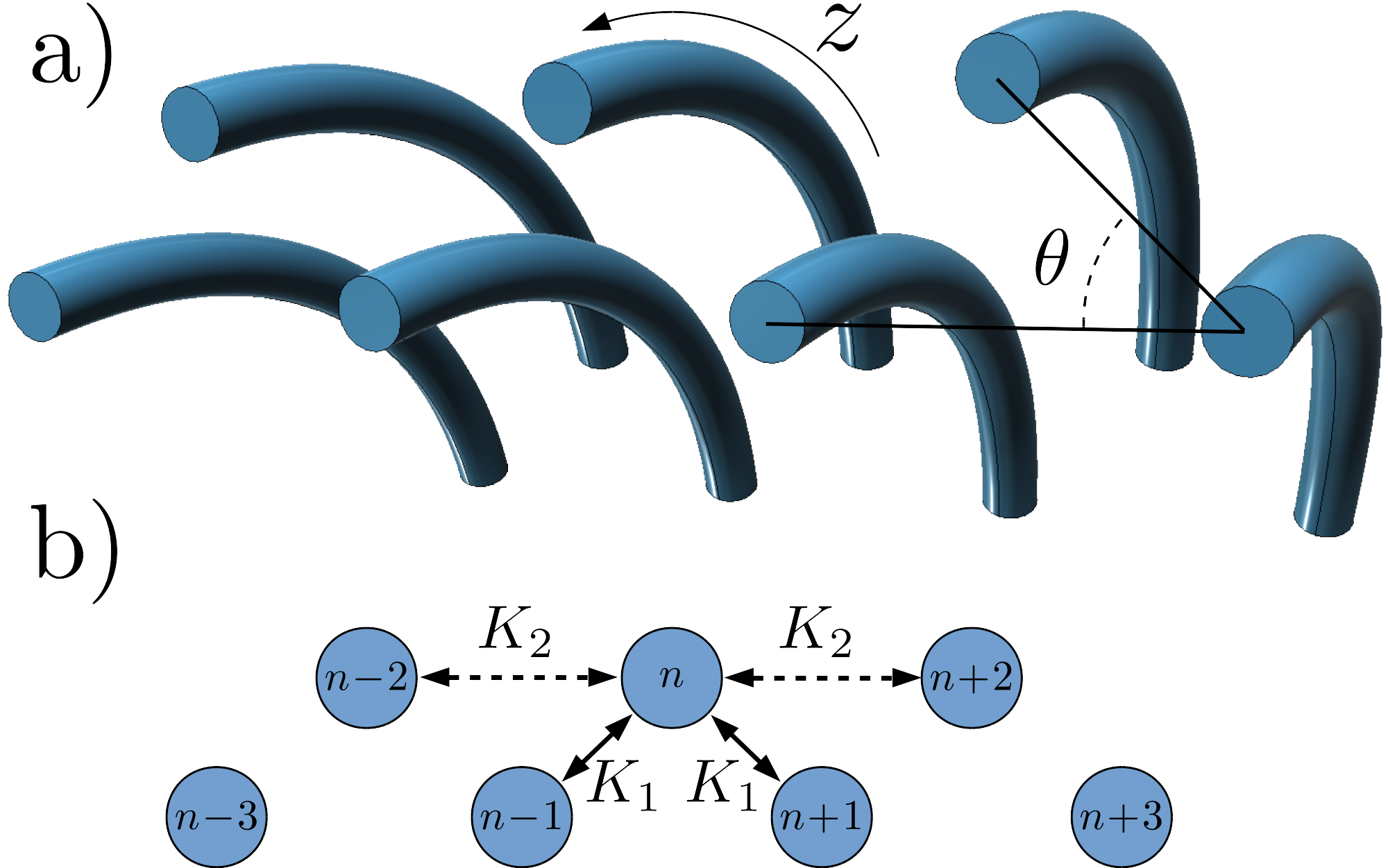}}
\caption{a)~Sketch of the zigzag arrays of curved waveguide reported in Ref.~\cite{Dreisow11}. b)~The interaction of magnitude $K_1$ between adjacent waveguides in the same row is indicated by the black solid arrow, whereas the second-order coupling $K_2$ is depicted by a black dashed arrow. The lattice angle $\theta$ determines the ratio $K_2/K_1$.}
\label{fig1}
\end{figure}

Firstly, we consider the light evolution in zigzag arrays of curved waveguides as those studied in Refs.~\cite{Wang10,Dreisow11} and shown schematically in Fig.~\ref{fig1}. The coupled mode equations for the field envelope $\Psi_n(z)$ in the $n$-th waveguide is
\begin{align}
i\,\frac{\mathrm{d}\Psi_{n}}{\mathrm{d}z}&-
\mathcal{F}n\Psi_{n}+K_1\left(\Psi_{n+1}+\Psi_{n-1}\right)\nonumber \\
&+K_2\left(\Psi_{n+2}+\Psi_{n-2}\right)=0\ ,
\label{eq:01}
\end{align}
where $z$ is the coordinate along the propagation direction, and $K_1$ and $K_2$ are the first- and second-order coupling constants, respectively. The transverse potential gradient $\mathcal{F}$ is inversely proportional to the curvature radius of the waveguides~\cite{Lenz99}. Defining $Z=K_1z$, $F=\mathcal{F}/K_1$ and $\beta_{2}=K_2/K_1$, Eq.~\eref{eq:01} can be cast in the form
\begin{align}
i\dot{\Psi}_n&-Fn\Psi_{n}+\Psi_{n+1}+\Psi_{n-1}\nonumber \\
&+\beta_{2}\left(\Psi_{n+2}+\Psi_{n-2}\right)=0\ ,
\label{eq:02}
\end{align}
where the dot indicates the derivative with respect to $Z$. Notice that the lattice angle $\theta$ in Fig.~\ref{fig1} determines the magnitude of $\beta_2$, as discussed in Ref.~\cite{Dreisow11}.

We now look for steady-state solutions of the form $\Psi_n(Z)=\psi_n\,\exp(-i\epsilon Z)$ satisfying
\begin{align}
\epsilon\psi_n&-Fn\psi_{n}+\psi_{n+1}+\psi_{n-1}\nonumber \\
&+\beta_{2}\left(\psi_{n+2}+\psi_{n-2}\right)=0\ .
\label{eq:03}
\end{align}
Although this problem can be tackled in Fourier space (see Ref.~\cite{Gozman16} for details), we shall follow a more convenient approach by expanding the amplitudes $\psi_n$ in the basis of solutions when $\beta_{2}=0$ instead of plane waves. In this way we are able to obtain closed expression for the amplitudes without relying on integrations in Fourier space. Additionally, we will show later that the results can be easily extended to cope with arrays subject to arbitrary long-range coupling. In the absence of second-order coupling ($\beta_{2}=0$), normalized solutions to~\eref{eq:03} in an infinite array can be expressed in terms of Bessel functions of integer order $J_{n-k}(2/F)$~\cite{Fukuyama73,Adame10}, where $k=0,\pm 1,\pm 2,\cdots$ labels the eigenstate. The corresponding eigenvalues are found to be $\epsilon_k=Fk$. Therefore, eigenvalues are equally spaced, also known as the Wannier-Stark ladder. 

For finite values of $\beta_{2}$ we propose the following combination of eigenstates for arrays with $\beta_{2}=0$ as a solution to~\eqref{eq:03}
\begin{equation}
{\psi_n}=\sum_{k=-\infty}^{\infty}c_k J_{n-k}(2/F) \ .
\label{eq:04}
\end{equation}
Using the orthonormality and the recurrence relations of the Bessel functions~\cite{Abramowitz72} one gets
\begin{equation}
(\epsilon-Fk)c_k+\beta_{2}(c_{k+2}+c_{k-2})=0\ .
\label{eq:05}
\end{equation}
Equation~\eqref{eq:05} relates coefficients $c_k$ with $k$ index of the same parity. Thus, the coefficients $c_k$ can be obtained by studying independently even and odd values of $k$. We can map~\eref{eq:05} onto a lattice problem with only first-order coupling by defining $\ell=(k-s)/2$, where $s=0$ for $k$ even and $s=1$ for $k$ odd. Hence
\begin{align}
&\big(\epsilon-Fs-2F\ell\big)c_{2\ell+s}\nonumber \\
&+\beta_{2}\left[c_{2(\ell+1)+s}+c_{2(\ell-1)+s}\right]=0\ .
\label{eq:06}
\end{align}
Solutions to~\eqref{eq:06} can be expressed in terms of Bessel functions of integer order and argument $\beta_{2}/F$. Finally, from~\eref{eq:04} we get
\begin{subequations}
\begin{equation}
{\psi_n}(q,s)=\sum_{k=-\infty}^{\infty} J_{k-q}(\beta_{2}/F)J_{n-2k-s}(2/F) \ ,
\label{eq:07a}
\end{equation}
for the normalized amplitudes and
\begin{equation}
\epsilon_{q,s}= (2q+s)F\ ,
\label{eq:07b}
\end{equation}
\label{eq:07}
\end{subequations}
for the nondegenerate eigenvalues, where $q=0,\pm 1,\pm 2,\cdots$ and $s=0,1$. Remarkably, eigenvalues are still equally spaced with spacing $F$, independently of the second-order coupling. 

\section{Wave packets under second-order coupling}   \label{sec:wave_packets}

In this section we focus on the dynamics of wave packets obeying Eq.~\eref{eq:02}, obtained as a combination of eigenstates~\eref{eq:07a} of the form
\begin{subequations}
\begin{equation}
\Psi_n(Z)=\sum_{q=-\infty}^{\infty}\sum_{s=0}^{1}\alpha_{q,s}\psi_{n}(q,s)e^{-i\epsilon_{q,s}Z}\ ,
\label{eq:08a}
\end{equation}
where the coefficients $\alpha_{q,s}$ are obtained from the initial wave packet as follows
\begin{equation}
\alpha_{q,s}=\sum_{n=-\infty}^{\infty}\psi_{n}(q,s)\Psi_{n}(0)\ .
\label{eq:08b}
\end{equation}
\label{eq:08}
\end{subequations}

We now calculate the centroid position $\xi(Z)=\langle n\rangle (Z) -\langle n\rangle(0)$ with
$\langle n\rangle(Z)=\sum_{n}n\left|\Psi_{n}(Z)\right|^2$. From Eq.~\eref{eq:08a} we get
\begin{equation}
\langle n\rangle(Z)
=\sum_{q,q^{\prime}}\sum_{s,s^{\prime}}\alpha_{q,s}^{*}
\alpha_{q^{\prime},s^{\prime}}\mathcal{S}_{qq^{\prime},ss^{\prime}}
e^{i(\epsilon_{q,s}-\epsilon_{q^{\prime},s^{\prime}})Z}\ ,
\label{eq:09a}
\end{equation}
where for brevity we have defined
$$
\mathcal{S}_{qq^{\prime},ss^{\prime}}\!=\!\sum_{k,k^{\prime}}J_{k-q}(\beta_{2}/F)
J_{k^{\prime}-q^{\prime}}(\beta_{2}/F)Q_{kk^{\prime},ss^{\prime}}\ ,
$$
with
$$
Q_{kk^{\prime},ss^{\prime}}=\sum_{n=-\infty}^{\infty}nJ_{n-2k-s}(2/F)J_{n-2k^{\prime}-s^{\prime}}(2/F)\ .
$$
The later summation is easily performed and the result is (see Appendix of Ref.~\cite{Adame10})
\begin{align}
&Q_{kk^{\prime},ss^{\prime}}=(2k+s)\delta_{2k+s,2k^{\prime}+s^{\prime}}\nonumber\\
&+\frac{1}{F}\,
\left(\delta_{2k+s,2k^{\prime}+s^{\prime}+1}+\delta_{2k+s,2k^{\prime}+s^{\prime}-1}\right)\ ,
\label{eq:10}
\end{align}
yielding
\begin{align}
&\mathcal{S}_{qq^{\prime},ss^{\prime}}\!=\!
\left[ (2q+s)\delta_{q,q^{\prime}}+\frac{\beta_{2}}{F}\left(\delta_{q,q^{\prime}+1}+\delta_{q,q^{\prime}-1}\right)
\right]\delta_{s,s^{\prime}} \nonumber \\
&+\frac{1}{F}\left[\delta_{q,q^{\prime}+(s^{\prime}-s+1)/2}+\delta_{q,q^{\prime}+(s^{\prime}-s-1)/2}\right]
\left(1-\delta_{s,s^{\prime}}\right)\ .
\label{eq:11}
\end{align}
Once $\mathcal{S}_{qq^{\prime},ss^{\prime}}$ is known, after some straightforward algebra we obtain the following closed expression for the centroid position
\begin{subequations}
\begin{equation}
\xi(Z)\!=\!\frac{2}{F}\,\mathrm{Re}\bigg[\left(e^{iFZ}-1\right)\mathcal{I}_1
+\beta_{2}\left(e^{2iFZ}-1\right)\mathcal{I}_2\bigg]
\label{eq:13a}
\end{equation}
%
where
\begin{equation}
\mathcal{I}_m=\sum_{n=-\infty}^{\infty}\Psi_{n}(0)\Psi_{n+m}^{*}(0)\ .
\label{eq:13b}
\end{equation}
\label{eq:13}
\end{subequations}
When $\beta_{2}=0$ we recover the results obtained in Ref.~\cite{Adame10} for a lattice with first-order coupling only, where the dynamics is characterized by the Bloch frequency $\omega_B=F$ with the chosen units, in agreement with the standard semiclassical calculation~\cite{Zener34}. However, at finite values of $\beta_{2}$ a frequency doubling sets up, as seen in~\eref{eq:13a}.

\begin{figure}[thb]
\includegraphics[width=\linewidth,clip]{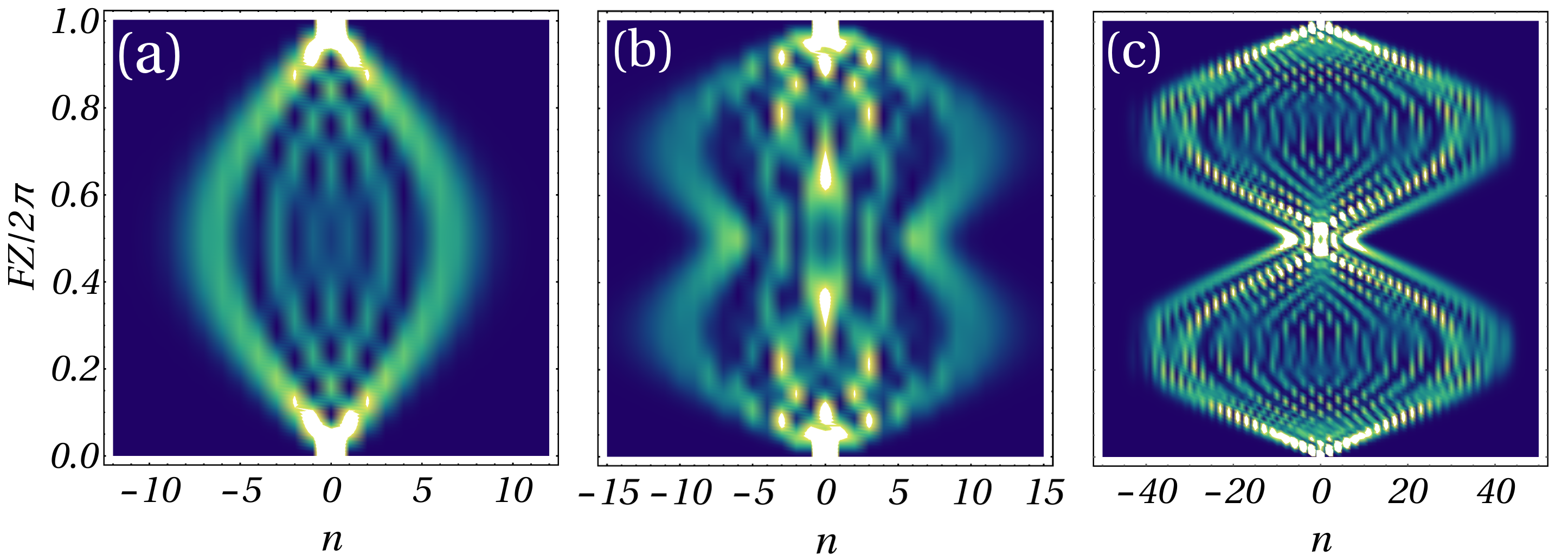}
\caption{Optical signal after excitation of a single waveguide at $n=0$, as a function of the propagation distance $Z$ and the the waveguide index $n$. Second-order couplings are a)~$\beta_2=0$, b)~$\beta_2=1.0$ and c)~$\beta=5.0$.}
\label{figbreathing}
\end{figure}

It is worth mentioning that we made no assumptions regarding the initial wave packet. In particular, Eq.~\eref{eq:13} is valid for broad as well as for narrow wave packets. We now prove that the light evolution strongly depends on the spatial width of the initial wave packet $\Psi_n(0)$. As working examples, let us consider two different initial conditions. In the first place, we assume that the wave-packet is initially localized at the origin. In this case $\Psi_{n}(0)=\delta_{n0}$ and the centroid vanishes at all times, as deduced from~\eref{eq:13} since $\mathcal{I}_1=\mathcal{I}_2=0$. Therefore, the wave packet remains at rest but develops a breathing mode, as already reported in Refs.~\cite{Dreisow11,Adame10,Harmann04} for a lattice with first-order coupling only. Breathing modes of the wave packet in zigzag arrays of curved waveguides with second-order coupling were already observed by Dreisow \emph{et al.} when only a single waveguide was excited ~\cite{Dreisow11}. 
Figure~\ref{figbreathing} shows the exact breathing dynamics for different values of the second-order coupling in agreement with experiments.

As a second example, let us discuss the other extreme, a very broad Gaussian wave packet initially at $Z=0$. When the width of the wave packet in units of the lattice spacing is large, the summations $\mathcal{I}_m$ defined by~\eref{eq:13b} equal approximately unity since $\Psi_{n+2}(0) \simeq \Psi_{n+1}(0) \simeq \Psi_{n}(0)$. Recalling that the semiclassical amplitude of the BOs is given as $A_B=2/F$~\cite{Adame10}, we get for the centroid
\begin{equation}
\xi(Z)\!=\!A_B\big\{\!\cos(\omega_BZ)-1
+\beta_{2}[\cos(2\omega_BZ)-1]\big\}\ ,
\label{eq:14}
\end{equation}
%
valid for a very broad wave packet only. A similar equation was found by Gozman \emph{et al.} for the optical signal path in arrays of coupled waveguides after solving the problem in Fourier space~\cite{Gozman16}. 

\begin{figure}[thb]
\centerline{\includegraphics[width=0.75\linewidth]{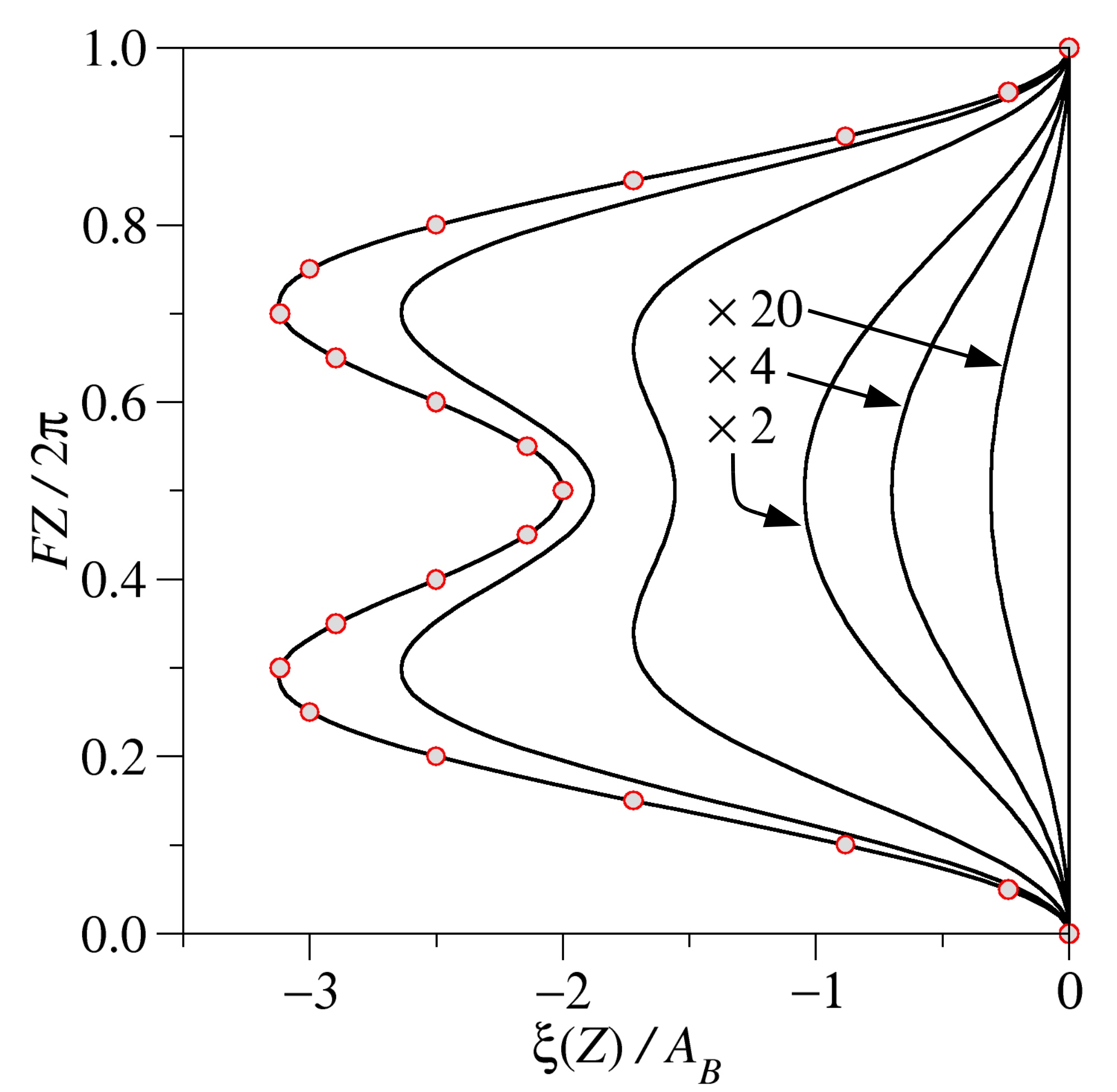}}
\caption{Crossover from broad to narrow wave packets, as depicted from the evolution of the centroid over a Bloch period. The initial wave packet is a normalized Gaussian with $\sigma=50.0$, $2.0$, $1.0$, $0.5$, $0.4$ and $0.3$ (from left to right) and $\beta_{2}=1$. Circles indicate the results from equation~\eref{eq:14}, only valid for very broad wave packets. Notice the scaling factor of the curves corresponding to the three lowest values of $\sigma$.}
\label{fig2}
\end{figure}

To study the crossover from narrow to broad wave packets, we consider an initial Gaussian wave packet $\Psi_{n}(0)=N\exp\left(-n^2/2\sigma^2\right)$, $N$ being a normalization constant. Figure~\ref{fig2} shows the evolution of the centroid $\xi(Z)$, given by Eq.~\eref{eq:13}, in units of the Bloch amplitude $A_B$, over a Bloch period. We set $\beta_{2}=1$ in the calculations. The anomalous BO due to the second-order coupling reveals itself by the presence of a local minimum at $\omega_BZ=0.5$. This minimum is clearly observed even if the width of the wave packet is of the order of the next-nearest neighbor distance ($\sigma=1.0-2.0$). The centroid explores a much smaller region of the system upon further decreasing the width of the initial wave packet ($\sigma < 1$), as expected. In this case the curves overlap with those obtained at $\beta_{2}=0$ (not shown in the figure) since the wave packet is strongly localized and cannot be influenced by the second-order coupling. From a more quantitative point of view, the product $\Psi_{n}(0)\Psi_{n+2}^{*}(0)$ is vanishingly small for narrow wave packets with $\sigma<1$ and the contribution of the last term in Eq.~\eref{eq:13a} can be neglected.

\section{Wave packets under long-range coupling}   \label{sec:long-range}

From simple geometrical considerations it is not difficult to see that the relative importance of the second-order coupling with respect to the first-order coupling increases upon increasing the lattice angle~$\theta$ (see Fig.~\ref{fig1}). But in this limit the magnitude of high-order couplings also increases. For instance, the largest angle considered in the experiments reported in Ref.~\cite{Dreisow11} is $\ang{75}$. At this value the third-order coupling is of the same order than the first-order coupling. While at large angles the light evolution is largely dominated by the second-order coupling, it becomes apparent that a more accurate description of the experiments is achieved by considering the effects of long-range coupling.

\begin{figure}[thb]
\centerline{\includegraphics[width=0.75\linewidth]{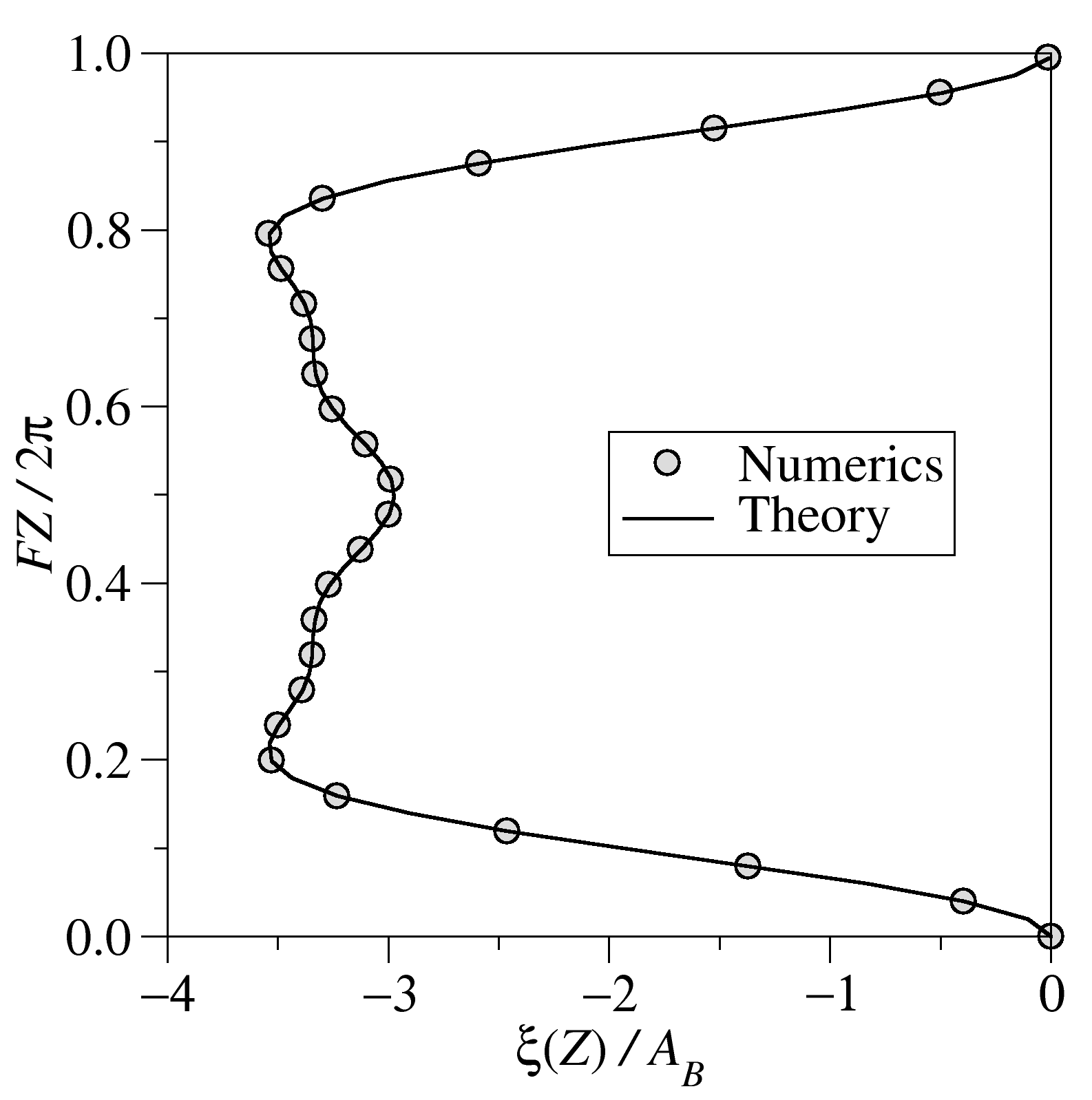}}
\caption{Evolution of the centroid over a Bloch period under long-range coupling with $N=4$. The initial wave packet is a normalized Gaussian with $\sigma=10.0$, and $\beta_{2}=1$, $\beta_{3}=0.5$ and $\beta_{4}=0.25$.}
\label{fig3}
\end{figure}

By inspecting Eq.~\eref{eq:13} it seems feasible to generalize the results to optical waveguide arrays with long-range coupling by repeating the renormalization procedure introduced in this work. Consider the dynamical problem in waveguide arrays with coupling beyond second-order. The coupled mode equations now read
\begin{equation}
i\dot{\Psi}_n-Fn\Psi_{n}+\sum_{m=1}^{N}\beta_{m}\left(\Psi_{n+m}+\Psi_{n-m}\right)=0\ ,
\label{eq:15}
\end{equation}
where $\beta_1\equiv 1$ and $N$ being an arbitrary positive integer. We do not attempt to solve the above equation but claim that the centroid of an arbitrary wave packet is given by the direct generalization of Eq.~\eref{eq:13a} as follows
\begin{equation}
\xi(Z)=\frac{2}{F}\,\mathrm{Re}\left[\sum_{m=1}^{N}\beta_{m}
\left(e^{imFZ}-1\right)\mathcal{I}_m\right]\ .
\label{eq:16}
\end{equation}
To validate the correctness of this statement we have solved Eq.~\eref{eq:15} numerically and obtained the dynamics of the centroid for different sets of parameters. Figure~\ref{fig3} shows the evolution of the centroid for an initial Gaussian wave packet with $\sigma=10.0$ subject to long-range coupling ($N=4$) and $\beta_{2}=1$, $\beta_{3}=0.5$ and $\beta_{4}=0.25$. Such scenario with decreasing couplings as a function of the index of neighboring waveguides is taken place when $\theta<\ang{60}$. 
Circles correspond to the numerical solution of Eq.~\eref{eq:15} and the solid line displays the evolution of the centroid as given by Eq.~\eref{eq:16}. Similar results were obtained for other sets of parameters. The agreement was always excellent and thus we claim that Eq.~\eref{eq:16} provides the correct evolution of the centroid of a wave packet in waveguide arrays subject to arbitrary long-range coupling. Notice that multiples of the Bloch frequency $\omega_B=F$ arise up to the order of the interaction. In a different context, Kosevich and Gann have demonstrated the appearance of higher harmonics of the Bloch frequency in the breathing dynamics of magnons in a spin chain with first-order coupling only, when they are prepared in a quantum superposition of non-nearest-neighbour Wannier-Zeeman states~\cite{Kosevich13}.

\section{Conclusions}

In this work we considered firstly the anomalous BOs that appear in zigzag arrays of curved waveguides when second-order coupling is significant. We expanded the steady state solutions in the basis of the eigenfunctions of the unperturbed problem, i.e. when only first-order coupling is relevant ($\beta_{2}=0$). The resulting problem was mapped onto a solvable one with first-order coupling due to the renormalization of the second-order coupling. Hence steady state solutions were expressed in terms of Bessel functions of integer order. The eigenvalues arrange into the well-known Wannier-Stark ladder, no matter the value of the second-order coupling. The dynamics of the wave packet was also solved exactly using those steady states. The centroid of the wave packet shows the frequency doubling phenomenon except if the initial wave packet is localized at a single site. In the latter case only breathing modes can be excited in the system. The crossover from broad to narrow wave packets was also discussed. Finally, we generalized our results to the case of arbitrary long-range coupling and found that multiples of the fundamental Bloch frequency, up to the order of the interaction, govern the dynamics of the wave packet in the array. 


\ack

The authors are indebted to A. D\'{\i}az-Fern\'{a}ndez and P. Hidalgo for helpful comments. This work was supported by the Spanish MINECO under Grants MAT2013-46308 and MAT2016-75955.

\section*{References}

\bibliography{references}

\bibliographystyle{iopart-num.bst}

\end{document}